\documentclass[useAMS,usenatbib]{mn2e}

\usepackage{amssymb}
\usepackage{times}
\usepackage{graphicx}
\usepackage{epstopdf}
\usepackage{subfigure}
\usepackage[T1]{fontenc}
\usepackage{aecompl} 
\usepackage{multirow}
\usepackage{pdflscape}
\usepackage{rotating}
\usepackage{hyperref}

\newcommand{\beq}{\begin{equation}}
\newcommand{\eeq}{\end{equation}}

\def\gs{\mathrel{\lower0.6ex\hbox{$\buildrel {\textstyle >}\over{\scriptstyle \sim}$}}}
\def\ls{\mathrel{\lower0.6ex\hbox{$\buildrel {\textstyle <}\over{\scriptstyle \sim}$}}}
\newcommand{\simgt}{\lower.5ex\hbox{$\; \buildrel > \over \sim \;$}}
\newcommand{\simlt}{\lower.5ex\hbox{$\; \buildrel < \over \sim \;$}}

\newcommand{\aap}{A\&A}
\newcommand{\apj}{ApJ}
\newcommand{\apjl}{ApJ}
\newcommand{\apjs}{ApJS}
\newcommand{\aj}{AJ}

\newcommand{\pasj}{PASJ}

\newcommand{\mnras}{MNRAS}
\newcommand{\physrep}{Phisycs Rep.}

\newcommand{\ssr}{Space Science Reviews}

\begin{document}

\title[LC$^2$]{CoMaLit -- III. Literature Catalogs of weak Lensing Clusters of galaxies (LC$^2$)}
\author[
M. Sereno
]{
Mauro Sereno$^{1}$\thanks{E-mail: mauro.sereno@unibo.it (MS)}
\\
$^1$Dipartimento di Fisica e Astronomia, Universit\`a di Bologna, viale Berti Pichat 6/2, 40127 Bologna, Italia\\
}


\maketitle

\begin{abstract}
The measurement of the mass of clusters of galaxies is crucial for their use in cosmology and astrophysics. Masses can be efficiently determined with weak lensing (WL) analyses. I compiled Literature Catalogs of weak Lensing Clusters (LC$^2$). Cluster identifiers,  coordinates, and redshifts have been standardised. WL masses were reported to over-densities of 2500, 500, 200, and to the virial one in the reference $\Lambda$CDM model. Duplicate entries were carefully handled. I produced three catalogs: LC$^2$-{\it single}, with 485 unique groups and clusters analysed with the single-halo model; LC$^2$-{\it substructure}, listing substructures in complex systems; LC$^2$--{\it all}, listing all the 822 WL masses found in literature. The catalogs and future updates are publicly available at \url{http://pico.bo.astro.it/\textasciitilde sereno/CoMaLit/LC2/}.
\end{abstract}


\section{Introduction}

Clusters of galaxies are at the crossroad between cosmology and astrophysics. They are laboratories to study the physics of the baryons and of the dark matter at large scales in bound objects \citep{voi05,pra+al09,arn+al10,gio+al13}. Cosmological parameters can be measured with cluster abundances and the observed growth of massive galaxy clusters \citep{man+al10,planck_2013_XX}, with gas fractions \citep{ett+al09}, or lensing analyses \citep{ser02,jul+al10,lub+al13}. This requires precise and accurate measurements of the cluster masses.

Cluster properties that can be easily measured with ongoing and future large surveys \citep{eucl_lau_11}, such as optical richness, X-ray luminosity and Sunyaev-Zel'dovich (SZ) flux, are going to be used as mass proxies. This relies on an accurate calibration through comparison with direct mass estimates \citep{an+be12,ett13,se+et14,ser+al14}. 

Weak lensing (WL) analyses provide one of the most well regarded mass estimate \citep{ba+sc01}. The physics behind gravitational lensing is well understood. The shear distortions of the background galaxies trace the gravitational field of the matter distribution of the lens \citep{hoe+al12,wtg_I_14,ume+al14}.  

Even if the WL estimate of the total projected mass along the line of sight is precise, the approximations that have to be used (spherical symmetry, smooth density distributions, no other contribution along the line of sight) to infer the three-dimensional mass may bias and scatter the results.

The main sources of uncertainty in WL mass estimates are due to triaxiality and substructures. The spherical assumption can bias the results for triaxial clusters pointing towards the observer, wherein lensing strengths are boosted and mass and concentration are over-estimated, or for clusters elongated in the plane of the sky, wherein mass and concentration are on the contrary under-estimated \citep{ogu+al05,ser07,cor+al09,ser+al10a,se+um11,se+zi12}.

Substructures in the cluster surroundings may dilute the tangential shear signal \citep{men+al10,gio+al12a,gio+al14}. Significant mass under-estimations can be caused by either massive sub-clumps just outside the virial radius \citep{men+al10} or uncorrelated large-scale matter projections along the line of sight \citep{be+kr11}.

Numerical studies have quantified the extent to which bias and intrinsic scatter affect WL masses. Usual fitting procedures of the cluster tangential shear profiles can bias low the mass by $\sim$5-10 per cent with a scatter of $\sim$10-25 per cent \citep{men+al10,be+kr11,ras+al12}. The exact value of the bias depends both on cluster mass and on radial survey range \citep{bah+al12}. The scatter should be less significant in optimally selected clusters either having regular morphology or living in substructure-poor environments \citep{ras+al12}. 

These theoretical predictions agree with recent measurements. \citet{se+et14} determined an intrinsic scatter for WL masses of $\sim$15 per cent. The scatter was estimated by comparing WL to X-ray masses based on the hypothesis of hydrostatic equilibrium in a number of well observed clusters from either the CLASH  \citep[Cluster Lensing And Supernova survey with Hubble,][]{pos+al12,ume+al14}, the CCCP  \citep[Canadian Cluster Comparison Project,][]{hoe+al12,mah+al13}, or the WtG \citep[Weighing the Giants,][]{wtg_I_14,wtg_III_14} programs.

An alternative and popular method to infer the cluster mass is based on the assumption that hydrostatic equilibrium holds between the intra-cluster medium (ICM) and the gravitational potential. The cluster mass can then be recovered from observations of the X-ray temperature and surface brightness \citep{lar+al06,don+al14}. However, deviations from equilibrium or non-thermal contributions to the pressure are difficult to quantify and can bias the mass estimate to a larger extent than for WL masses \citep{ras+al12,se+et14}.

Other methods to derive the cluster mass employ spectroscopic measurements of galaxies velocities, such as the caustic technique \citep{ri+da06}, or approaches exploiting the Jeans equation \citep{lem+al09,biv+al13}. These methods require observations more expansive than photometric surveys and are mostly limited to low redshift halos.

WL masses can be obtained up to high redshifts in the context of large photometric surveys, and they are nearly unbiased. They are supposedly the best mass estimators to calibrate other proxies.

In this paper I re-elaborate in a standard form known WL mass estimates of galaxy clusters available in literature. The typical information presented in WL studies is not standardised. A cluster can be named in different ways. Different conventions are employed for the reference cosmological model. The lens can be characterised in a number of ways. A quantitative analysis can provide either the total mass within an integration radius (which on turn can be defined in several ways), or the total projected mass within an angular aperture (this is the quantity the lensing is most sensitive to), or the parameters characterising the adopted mass profile. 

I collected all the disparate WL measurements available in literature in three meta-catalogs regularised to the same reference cosmology and to the same set of integration radii. The basic characteristics of these catalogues are the large number of objects (485 unique systems), and the standardised names, coordinates, redshifts and masses. References to the original analyses were reported for each cluster.

I compiled three catalogues: $i$) the LC$^2$-{\it single} lists the unique systems. Duplicate entries originating from overlaps between the input references were controlled and eliminated. The reported masses of either regular or complex clusters were obtained with a single-halo analysis. These are the most sensible masses to compare to other global properties, such as the SZ flux, the X-ray luminosity or the optical richness. $ii)$ The LC$^2$-{\it substructure} lists the main and the secondary substructures of complex clusters which were studied with a multiple-halo analysis. The mass of each component is reported individually. $iii)$ The LC$^2$-{\it all} lists all the groups and clusters found in literature. Repeated entries are included. LC$^2$-{\it single} and LC$^2$-{\it substructure} are subsamples of LC$^2$-{\it all}.

The catalogs are publicly available in electronic format and will be periodically updated. Updates can be found at \url{http://pico.bo.astro.it/~sereno/CoMaLit/LC2/}.

For the compilation of the catalogs, I assumed a fiducial flat $\Lambda$CDM cosmology with density parameter $\Omega_\mathrm{M0}=0.3$, and Hubble constant $H_0=70~\mathrm{km~s}^{-1}\mathrm{Mpc}^{-1}$. 

This paper is the third in a series titled `CoMaLit' (COmparing MAsses in LITerature). In the first paper \citep[ CoMaLit-I]{se+et14}, systematic differences in lensing and X-ray masses obtained from independent analyses were quantified and the overall level of bias and intrinsic scatter was assessed through Bayesian techniques. This formalism was later applied and developed in the second paper of the series \citep[ CoMaLit-II]{ser+al14} to calibrate the Sunyaev-Zel'dovich (SZ) flux estimated by the Planck satellite against mass proxies. The fourth paper \citep[ CoMaLit-IV]{se+et14b} studies the time-evolution of the scaling relations.

The papers is structured as follows. In Section~\ref{sec_meta}, I comment on qualities and drawbacks of meta-catalogues collected from literature and on their use in astronomy. In Section~\ref{sec_mass}, I review the various definitions of over-density and virial radii and I motivate the choice of the radii used for the catalogs. In Section~\ref{sec_prof}, I summarise the properties of the most used mass density distributions to characterise the lens and I discuss how I standardised the estimates of the masses listed in the catalogs. Section~\ref{sec_cosm} discusses the dependence of the WL mass estimates on the cosmological parameters and how they can be uniformed to a given reference cosmological model.  In Section~\ref{sec_cata}, I discuss how I assembled the catalogs from the various literature sources and  how I performed the cluster identifications. Section~\ref{sec_pres} is devoted to the presentation of the format of the catalogs. Final considerations are in Section~\ref{sec_summ}.

\section{On meta-catalogues}
\label{sec_meta}

The worthiness of coherently compiled meta-catalogues of clusters has been discussed in \citet{pif+al11}, who collected a large catalogue of X-ray detected clusters of galaxies based on publicly available samples.

Specifically to the WL catalogs here presented, I remark that the LC$^2$-{\it all} provides a panorama of the state-of-the-art on weak lensing clusters. It gives an overview of the published, publicly available weak lensing analyses. It is a repository of references and a ready-to use collection of the main properties (coordinates, redshift and mass) of the observed clusters.

Large, standardised catalogues can be used for cross-correlation with existing, ongoing, or upcoming surveys at various wave-lengths, such as SZ \citep{planck_2013_XX,rei+al13,men+al13}, optical \citep[ Euclid]{eucl_lau_11}, or X-ray surveys \citep[ and references therein]{pif+al11}. The mass, in combination with the appropriate scaling laws, enables us to predict all the main properties of the clusters, such as the integrated SZ flux, the X-ray temperature, the optical richness, and the velocity dispersion.

The largest public catalogs of massive WL clusters consists of a few dozens of objects \citep{sha+al12,mah+al13,wtg_III_14,ume+al14}. Clusters are not usually selected according to strict selection functions and some sort of arbitrariness can persist. The usual WL sample that can be found in literature is then small but it is neither statistical nor complete. It can be worthy to take a different route, i.e., to consider a sample whose selection function is not known but that is as large as possible. A very large sample, no matter whether it was assembled in a heterogeneous way, can recover the actual physical trends we are looking for \citep{got+al01}.

The LC$^2$ catalogues can be useful for the construction of better defined subsamples. The full sample of collected clusters is neither statistical nor complete. The reconstruction of the selection function of meta-catalogues is a nearly impossible task \citep{pif+al11}. The individual selection functions of the subsamples are complex and, in most cases, are not known or not available. However, suitable subsamples can be extracted for which the selection function can be approximated. These subsamples can be used to study scaling relations, time evolution of structures, and cosmography.

A large collection of clusters enables us to assess the reliability of the WL mass measurements \citep{se+et14}. The repeated entries in LC$^2$-{\it all}  can be used to compare mass estimates from different analyses. Published uncertainties are often unable to account for the actual variance seen in sample pairs \citep{roz+al14,se+et14}. The certain assessment of cluster masses is hindered by instrumental and methodological sources of errors which may cause systematic uncertainties in data analysis \citep{roz+al14}. The main sources of systematics in lensing analyses are due to selection and calibration problems. The selection and redshift measurement of background galaxies is a very difficult task that has to be undertaken through accurate photometric redshifts and colour-colour selection methods \citep{med+al10,gru+al14}. A small calibration correction of the shear signal of the order of just a few percents can produce a systematic error of $\sim$ 10 per cent in the estimate of the virial mass \citep{ume+al14}. Differences in WL mass estimates reported by different groups can be as large as $\sim$40 per cent \citep{se+et14}. 

Even though the catalogues are presented in a uniform format, I remark that they are highly heterogeneous. The clusters were detected in a variety of ways within X-ray, optical, SZ, or shear surveys. Some clusters were targeted because they are very peculiar objects, as merging \citep{ok+um08} or high-redshift clusters \citep{jee+al11}. Some samples of clusters were assembled based on their known properties, as their X-ray luminosity or regular X-ray morphology \citep{mah+al13,wtg_I_14,ume+al14}. Others were observed in follow-up programs of differently planned surveys, which significantly increased the number of studied lensing clusters and extended the observation range to lower mass objects \citep{ket+al13,mci+al09}. Some samples were shear selected \citep{sha+al12}.

On the positive end, systematic biases that affect some specific, small samples may average out in a heterogeneous and very large sample. The larger the sample, the smaller the biases due to the orientation of the clusters, to their internal structure, and to the projection effect of large-scale structure. Due to the different finding techniques, biases plaguing lensing selected samples, such as the over-concentration problem and the orientation bias \citep{og+bl09,men+al11}, are mitigated too. Projection effects are less severe in X-ray or SZ detected clusters.

The different observational facilities and data analysis methods also increase the heterogeneous nature of the catalog. Different solutions to instrumental and methodological sources of errors may cause systematic errors in the mass determination. The heterogeneity of the catalogs manifests both in the listed central estimates and the uncertainties. Masses are presented in a homogeneous way but they were not derived homogeneously among the original studies.

\section{Masses}
\label{sec_mass}

Total masses of clusters within an over-dense region can be related to the virial mass. Most cluster properties are expected to be self-similar at those scales. There are several commonly used definitions of the virial radius. Over-densities can be measured either with respect to the critical density of the universe at the epoch of analysis,  ($\Delta_\mathrm{cr}$), or with respect to the mean density ($\Delta_\mathrm{m}$). For the compilation of the catalog, I considered $\Delta = \Delta_\mathrm{cr}$, in terms of which important properties of galaxy clusters are universal \citep{di+kr14}.

$M_{\Delta}$ denotes the mass within the radius $r_{\Delta}$, which encloses a mean over-density of $\Delta$ times the critical density at the cluster redshift, $\rho_\mathrm{cr}=3H(z)^2/(8\pi G)$; $H(z)$ is the redshift dependent Hubble parameter. By definition, $M_\Delta$ can be expressed as
\beq
\label{eq_over_1}
M_\Delta =\frac{4\pi}{3}\Delta \rho_\mathrm{cr} r_\Delta^3.
\eeq

Numerical simulations showed that fixed over-densities are very useful to describe universal features of clusters and to study the scaling relations \citep{tin+al08,di+kr14}. From the theoretical point of view, the virialised region of a cluster can be related to the solution to the collapse of top-hat perturbations.  The viral over-density is then redshift and cosmology dependent. To compute the virial radius, I adopted the approximated relation proposed by \citet{br+no98}, which is based on the spherical collapse model for a flat universe with cosmological constant,
\beq
\label{eq_over_1b}
\Delta_\mathrm{vir} \simeq 18\pi^2+82 [\Omega_\mathrm{M}(z)-1]-39 [\Omega_\mathrm{M}(z)-1]^2.
\eeq

WL studies probe the clusters on large radial scales. As integration radii, I considered the virial radius and $r_{200}$, which usually enclose most of the field of view covered by observations and are also well probed by SZ analyses; $r_{500}$, which still encloses a substantial fraction of the total virialised mass of the system and is usually the largest radius probed in X-ray observations; $r_{2500}$, which is usually poorly constrained by WL alone, but that can still be useful in comparison with detailed analysis of the cluster core, as those based on current high resolution X-ray observations or strong lensing investigations. Results at $r_{2500}$ are mostly based on extrapolations and they may be unreliable without strong lensing constraints. 

The critical surface density for lensing is defined as 
\beq
\Sigma_\mathrm{cr}\equiv\frac{c^2\,D_\mathrm{s}}{4\pi G\,D_\mathrm{d}\,D_\mathrm{ds}},
\eeq
where $D_\mathrm{s}$, $D_\mathrm{d}$ and $D_\mathrm{ds}$ are the source, the lens and the lens-source angular diameter distances, respectively.

\section{Mass profiles}
\label{sec_prof}

Whenever the masses $M_\Delta$ were quoted in the original papers, I took them for the catalogs. If not, I had to extrapolate the quoted results based on the density profile adopted in the analysis. The Navarro-Frenk-White profile \citep[ NFW]{nfw96}, and the singular isothermal sphere (SIS) are the standard parametric models used in lensing analyses to characterise the deflector. 

Alternatively, some works quote only the total projected mass in an angular aperture. This may be the case of combined strong and weak lensing analyses or of free-form modelling. In these cases, I extrapolated the results by adopting a NFW model.

\subsection{NFW}

Dark matter halos are successfully described as NFW density profiles \citep{nfw96,ji+su02}. The 3D density distribution follows
\begin{equation}
\label{eq_nfw_1}
	\rho_\mathrm{NFW}=\frac{\rho_\mathrm{s}}{(r/r_\mathrm{s})(1+r/r_\mathrm{s})^2},
\end{equation}
where $r_\mathrm{s}$ is the scale radius. The mass enclosed at radius $r$ is
\beq
\label{eq_nfw_2}
M_\mathrm{NFW}(<r)=4 \pi\ \rho_\mathrm{s}\ r_\mathrm{s}^3 F_\mathrm{NFW}(r_\mathrm{s}/r),
\eeq
where
\beq
\label{eq_nfw_3}
F_\mathrm{NFW}(x)=x^3 \left[ \ln(1+x^{-1}) - (1+x)^{-1}\right].
\eeq
The NFW model is characterised by two parameters. They can be $\rho_\mathrm{s}$ and $r_\mathrm{s}$, or the mass $M_\Delta$ and the concentration, $c_{\Delta} \equiv r_{\Delta}/ r_\mathrm{s}$. The conversion relations are simple. From the definition of concentration and Eq.~(\ref{eq_nfw_2}),
\beq
\label{eq_nfw_4a}
r_\mathrm{s} = r_{\Delta}/  c_{\Delta} 
\eeq
and
\beq
\label{eq_nfw_4b}
\rho_\mathrm{s} = \frac{\Delta}{3} \frac{1}{ F_\mathrm{NFW}(1/c_{\Delta} )} \rho_\mathrm{cr}.
\eeq


The general conversion from a mass at an arbitrary over-density, $\Delta_1$, to a second one, $\Delta_2$, was derived in \citet{hu+kr03}. By writing the parameters $r_\mathrm{s}$  and $\rho_\mathrm{s} $ in terms of two different over-densities through Eqs.~(\ref{eq_nfw_4a} and \ref{eq_nfw_4b}) and equating the expressions, we obtain
\begin{eqnarray}
F_\mathrm{NFW}\left(\frac{1}{c_{\Delta_2}}\right) & =& \frac{\Delta_2}{\Delta_1}F_\mathrm{NFW}\left(\frac{1}{c_{\Delta_1}}\right), \label{eq_nfw_5} \\
M_{\Delta_2}  & =& \frac{\Delta_2}{\Delta_1} \left( \frac{c_{\Delta_2}}{c_{\Delta_1}} \right)^3 M_{\Delta_1} . \label{eq_nfw_6}
\end{eqnarray}
The conversion involves the inversion of the function $F_\mathrm{NFW}(x)$.

Equations~(\ref{eq_nfw_5},~\ref{eq_nfw_6}) can also be rewritten to derive the concentrations given two integrated masses, $M_{\Delta_1}$ and $M_{\Delta_2}$,

\beq
F_\mathrm{NFW}\left( \frac{1}{c_{\Delta_1}} \left( \frac{\Delta_2 M_{\Delta_1}}{\Delta_1 M_{\Delta_2}} \right)^{1/3}  \right)  = \frac{\Delta_2}{\Delta_1}F_\mathrm{NFW}\left(\frac{1}{c_{\Delta_1}}\right); \label{eq_nfw_7} 
\eeq
given two masses and one concentration, the remaining concentration can be obtained as
\beq
c_{\Delta_2} =  c_{\Delta_1} \left( \frac{\Delta_1 M_{\Delta_2}}{\Delta_2 M_{\Delta_1}} \right)^{1/3}.
\eeq

An additional relation has to be used to constrain the NFW profile if only one parameter is known. $N$-body simulations have proved that mass and concentration are related \citep{net+al07,gao+al08,duf+al08,pra+al11,du+ma14,di+kr14}. In limited ranges, the dependence of the halo concentration on mass and redshift can be adequately described by a power law, 
\beq
\label{eq_cM_1}
c_{200} =A(M_{200}/M_\mathrm{pivot})^B(1+z)^C.
\eeq
Simulations shows that concentrations are scattered about the median relation. The scatter is approximately log-normal and it is of the order of $\sim 30$ per cent \citep{duf+al08,bha+al13}. Whereas different studies agree on the functional form of the relation \citep{di+kr14} and on the level of scatter, some disagreement as large as $\sim50$ per cent is still present on the overall normalisation of the relation \citep{pra+al11}. 

On the observational side, the concentration estimated with WL studies can differ from the intrinsic one due to uncorrelated or correlated large-scale structure, baryonic physics, and, mainly, triaxiality and orientation of the halo ellipsoid with respect to the line-of-sight \citep{bah+al12,gio+al14}. 

If only one parameter is reported in the analysis, I broke the degeneracy in the mass profile by adopting the relation in Eq.~(\ref{eq_cM_1}) with $A = 5.71$, $B= -0.084$, and $C= -0.47$ for a pivotal mass $M_\mathrm{pivot}=2\times10^{12}M_\odot/h$ \citep{duf+al08}. Uncertainties and scatter can affect the estimation of the extrapolated masses when the concentration is determined with a given mass-concentration relation. A deviation of the order of $\sim 30$ per cent from the median $c_{200}$ in a large mass ($10^{14}M_\odot\ls M_{200}\ls 10^{15}M_\odot$) and redshift ($z\ls 1$) range causes an analog deviation on the estimate of $M_{500}$ of the order of $\sim$20-30 per cent. On the other hand, the estimate of $M_{200}$ from the analysis of the shear profile depends weakly on the assumed concentration \citep{wtg_III_14}.

Some analyses quote only the projected mass within an aperture radius. The total projected mass for a NFW lens can be expressed as
\beq
\label{eq_nfw_9}
M^\mathrm{cyl}_\mathrm{NFW}(<R)=4\pi \rho_\mathrm{s} r_\mathrm{s}^3 \left\{  2\frac{\mathrm{arctanhh}\left| \frac{1-x}{1+x} \right|}{\sqrt{|1-x^2|}}  +\ln \left(\frac{x}{2}\right) \right\},
\eeq
where $x$ is the dimensionless projected radius, $x \equiv R /r_\mathrm{s}$, and $\mathrm{arctanhh} =\mathrm{arctanh}$ ($\arctan$) if $x<(>)1$.
If only the mass within a cylinder, $M^\mathrm{cyl}_\mathrm{obs}$, is provided, the mass $M_{\Delta}$ can be derived by inverting
\beq
M^\mathrm{cyl}_\mathrm{NFW} (R_\mathrm{obs}; M_{\Delta},c_\Delta (M_{\Delta}))  =  M^\mathrm{cyl}_\mathrm{obs}, \label{eq_nfw_10}
\eeq
where $c_\Delta (M_{\Delta})$ can be expressed as in Eq.~(\ref{eq_cM_1}).



\subsection{Singular Isothermal sphere}

An alternative mass profile is provided by the singular isothermal sphere \citep{tur+al84}, whose density profile is
\beq
\label{eq_sis_1}
\rho_\mathrm{SIS}= \frac{1}{2\pi}\frac{\sigma_\mathrm{SIS}^2}{G}\frac{1}{r^2}.
\eeq
This model was the standard for lens profiles before being supplanted by the NFW model. The total mass within a spherical radius is
\beq
\label{eq_sis_2}
M_\mathrm{SIS}(<r)= \frac{2 \sigma_\mathrm{SIS}^2}{G} r .
\eeq
It follows that
\beq
\label{eq_sis_3}
r_\Delta= \frac{2 \sigma_\mathrm{SIS}^2}{H(z)\sqrt{\Delta}},
\eeq
and
\beq
\label{eq_sis_4}
M_\Delta= \frac{4 \sigma_\mathrm{SIS}^3}{G H(z)\sqrt{\Delta}} .
\eeq

\section{Cosmological parameters}
\label{sec_cosm}

\begin{figure}
       \resizebox{\hsize}{!}{\includegraphics{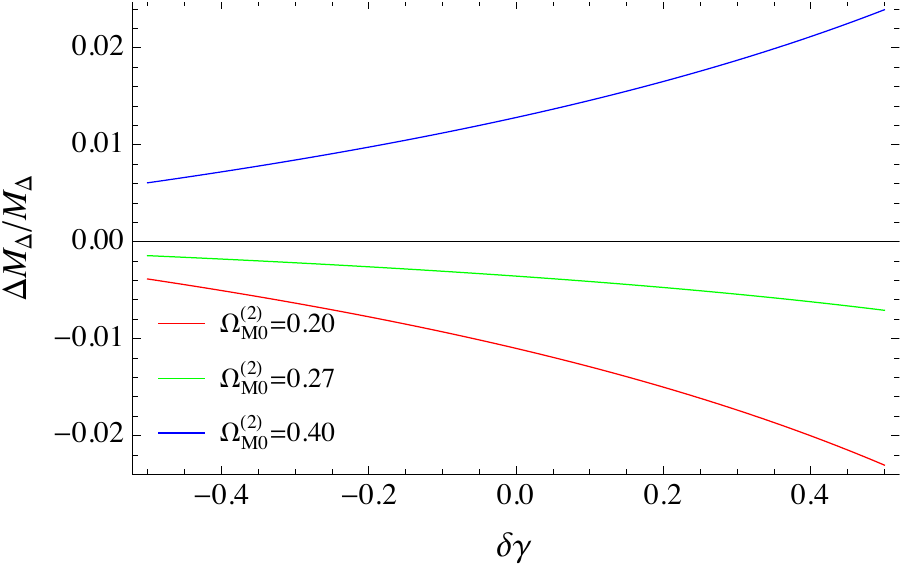}}
       \caption{Relative variation of the estimated WL mass within a fixed over-density, $M_{\Delta}$, as a function of the slope $\delta\gamma$ for different (flat) cosmological models with respect to the standard $\Lambda$CDM model with $\Omega_\mathrm{M0}=0.3$. The lens redshift is $z_\mathrm{d}=0.3$; the background galaxies are at $z_\mathrm{s}=1.0$. The red, green, and blue lines refer to flat $\Lambda$CDM models with $\Omega_\mathrm{M0}=$0.20, 0.27 and 0.40, respectively}
	\label{fig_DeltaM_deltagamma}
\end{figure}

Lensing mass estimates depend on the assumed cosmological model. If necessary, they were rescaled to the reference cosmological model, i.e., a flat $\Lambda$CDM cosmology with density parameter $\Omega_\mathrm{M0}=0.3$, and Hubble constant $H_0=70~\mathrm{km~s}^{-1}\mathrm{Mpc}^{-1}$. 

The lensing 3D mass within a radius $r=D_\mathrm{d}\theta$, where $\theta$ is the angular radius, scales as \citep{se+et14}
\beq
\label{eq_over_2}
M^\mathrm{WL} \propto \Sigma_\mathrm{cr} D_\mathrm{d}^2 \theta_\mathrm{E}~\theta f(\theta),
\eeq
where $ \theta_\mathrm{E}$ is the angular Einstein radius. The function $f(\theta)\sim\theta^{\delta\gamma}$ quantifies the deviation of the mass profile from the isothermal case. 

By equating Eq.~(\ref{eq_over_1}) and  Eq.~(\ref{eq_over_2}) at $\theta_\Delta (=r_\Delta/D_\mathrm{d})$, we obtain 
\beq
\label{eq_over_3}
M^\mathrm{WL}_\Delta   \propto  D_\mathrm{d}^{-\frac{3\delta\gamma}{2-\delta\gamma}}\left(   \frac{D_\mathrm{ds}}{D_\mathrm{s}} \right)^{-\frac{3}{2-\delta\gamma}} H(z)^{-\frac{1+\delta\gamma}{1-\delta\gamma/2}} .
\eeq

Equation~(\ref{eq_over_3}) holds for a fixed over-density, whereas the viral over-density depends on the cosmological parameters. For the virial mass,
\beq
\label{eq_over_5}
M^\mathrm{WL}_\mathrm{vir}   \propto \Delta_\mathrm{vir}^{-\frac{1+\delta\gamma}{2-\delta\gamma}} D_\mathrm{d}^{-\frac{3\delta\gamma}{2-\delta\gamma}}\left( \frac{D_\mathrm{ds}}{D_\mathrm{s}} \right)^{-\frac{3}{2-\delta\gamma}} H(z)^{-\frac{1+\delta\gamma}{1-\delta\gamma/2}} ,
\eeq
where $\Delta_\mathrm{vir}$ is a function of the redshift dependent cosmological density, see Eq.~(\ref{eq_over_1b}).

The dependence on the cosmological parameters is usually small. The variation is $\ls 2$ per cent for a large range of mass profiles and cosmological models, see Fig.~\ref{fig_DeltaM_deltagamma}. 

The condition $\delta\gamma=0$ is strictly verified only for the singular isothermal profile but it provides a good approximation in general. Let us consider as a typical massive lens, a NFW distribution with $M_{200} \simeq10^{15}M_\odot$ and $c_{200} \simeq 3$.  The deviation of the slope from the isothermal value is small over a large radial range, with $\delta\gamma (r_{2500}) \simeq 0.4$, $\delta\gamma (r_{500}) \simeq 0.0$, $\delta\gamma (r_{200}) \simeq -0.1$, $\delta\gamma (r_\mathrm{vir}) \simeq -0.2$.

To make the proper conversion from different cosmological parameters, I used by default $\delta\gamma=0$, when Eq.~(\ref{eq_over_3}) reduces to 
\beq
\label{eq_over_4}
M^\mathrm{WL}_\Delta   \propto  \left(  \frac{D_\mathrm{ds}}{D_\mathrm{s}} \right)^{-3/2}H(z)^{-1} .
\eeq
and Eq.~(\ref{eq_over_5}) can be simplified as
\beq
\label{eq_over_6}
M^\mathrm{WL}_\mathrm{vir}   \propto \Delta_\mathrm{vir}(\Omega_\mathrm{M})^{-\frac{1}{2}} \left(  \frac{D_\mathrm{ds}}{D_\mathrm{s}} \right)^{-3/2}H(z)^{-1} . 
\eeq

\section{Catalog compilation}
\label{sec_cata}

\begin{table}
\caption{Number of clusters, groups, or substructures ($N_\mathrm{clusters}$ in col.~3), analysed in each reference, col.~1. The authors' code is listed in col.~2.}
\label{tab_refe}
\centering
\resizebox{!}{11.5cm}{
\begin{tabular}{ l l r }     
Reference		 & Code &	 $N_\mathrm{clusters}$ \\ 
\hline
\citet{sha+al12}       & shan+12       & 87 \\
\citet{hoe+al12}       & hoekstra+12 & 55 \\
\citet{wtg_III_14}      &applegate+14   & 51 \\
\citet{mah+al13}        &mahdavi+13   & 50 \\
\citet{dah+al02}          &dahle+02   & 38 \\
\citet{mci+al09}           &mcinnes+09	& 36 \\
\citet{dah06}             &dahle06	& 35 \\
\citet{se+co13}          &sereno\&13  & 31 \\
\citet{oka+al10}          &okabe+10   & 30 \\
\citet{pe+da07}          &pedersen\&07	& 30 \\
\citet{ogu+al12}          &oguri+12   & 28 \\
\citet{ham+al09}         &hamana+09   & 27 \\
\citet{jee+al11}            &jee+11   & 27 \\
\citet{hoe+al11}          &hoekstra+11	& 25 \\
\citet{cyp+al04}          &cypriano+04	& 24 \\
\citet{clo+al06a}          &clowe+06  & 20 \\
\citet{ume+al14}          &umetsu+14  & 20 \\
\citet{mer+al14}            &merten+14	& 19 \\
\citet{gru+al14}             &gruen+14	& 12 \\
\citet{lim+al09}          &limousin+09& 12 \\
\citet{bar+al07}          &bardeau+07 & 11 \\
\citet{foe+al12}           &foex+12   & 11 \\
\citet{ket+al13}           &kettula+13	& 10 \\
\citet{sma+al97}          &smail+97   & 10 \\
\citet{aba+al09}            &abate+09 & 9  \\
\citet{ok+um08}             &okabe\&08		& 9  \\
\citet{ga+so07}           &gavazzi\&07		& 8  \\
\citet{isr+al12}            &israel+12		& 8  \\
\citet{kub+al09}           &kubo+09		   & 7  \\
\citet{wat+al11}          &watanabe+11	& 6  \\
\citet{clo+al00}            &clowe+00 & 6  \\
\citet{hig+al12}            &high+12  & 5  \\
\citet{ume+al11a}         &umetsu+11   & 5  \\
\citet{oka+al11}             &okabe+11	& 4  \\
\citet{ume+al09}            &umetsu+09	& 4  \\
\citet{mel+al14}          &melchior+14		& 4  \\
\citet{oka+al14b}         &okabe+14b	   & 4  \\
\citet{cor+al09}           &corless+09	& 3  \\
\citet{gra+al02}           &gray+02   & 3  \\
\citet{gav+al04}           &gavazzi+04& 3  \\
\citet{jee+al14}             &jee+14  & 3  \\
\citet{bra+al06}            &bradac+06	& 2  \\
\citet{ham+al12}   	&hamilton-morris+12	& 2  \\
\citet{die+al09}          &dietrich+09	& 2  \\
\citet{bra+al08b}        &bradac+08b   & 2  \\
\citet{bra+al08a} 	&bradac+08a	& 1 \\
\citet{clo+al04}  	&clowe+04	& 1 \\
\citet{gav05}     	&gavazzi05	& 1 \\
\citet{gav+al09}  	&gavazzi+09	& 1 \\
\citet{hal+al06}  	&halkola+06	& 1 \\
\citet{hic+al07}  &hicks+07	& 1 \\
\citet{hua+al11} &huang+11	 & 1 \\
\citet{jau+al12}  &jauzac+12	& 1 \\
\citet{jau+al14}  &jauzac+14& 1 \\
\citet{kub+al07}  &kubo+07	& 1 \\
\citet{ler+al11}  &lerchster+11	& 1 \\
\citet{lim+al07}  &limousin+07	& 1 \\
\citet{lim+al10}  &limousin+10	& 1 \\
\citet{mah+al07} &mahdavi+07	 & 1 \\
\citet{mar+al05}  &margoniner+05	& 1 \\
\citet{mer+al11}  &merten+11	& 1 \\
\citet{miy+al13}  & miyatake+13	&1 \\
\citet{ogu+al13}  & oguri+13	&1 \\
\citet{oka+al14a}&okabe+14a & 1 \\
\citet{pau+al07}  & paulin-henriksson+07	&1 \\
\citet{rad+al08}  &radovich+08	& 1 \\
\citet{rom+al10}  &romano+10	& 1 \\
\citet{sch+al10}  &schirmer+10	& 1 \\
\citet{sch+al11}  &schirmer+11	& 1  \\
\hline	
\end{tabular}
}
\end{table}


I included in the catalog all groups and clusters with weak lensing analyses I was aware of. The research in literature was performed thanks to the NASA's Astrophysics Data System\footnote{\url{http://www.adsabs.harvard.edu/}.}. A public list of clusters with weak lensing analyses, compiled by H.~Dahle and last updated in 2007, was also used\footnote{\url{http://folk.uio.no/hdahle/WLclusters.html}.}. 

The compilation of the first versions of the catalogs was based on 69 weak lensing studies comprising 822 analyses of individual groups and clusters, see Table~\ref{tab_refe}. 

The catalogs were meant to avoid re-elaboration as much as possible. Masses quoted in the reference papers were directly reported. When original estimates were provided with asymmetric errors, I computed the mean value and the standard deviation as suggested in \citet{dag04}. Missing masses were computed by extrapolation as discussed in Sec.~\ref{sec_prof}. Corrections for the cosmological model were performed as detailed in Sec.~\ref{sec_cosm}.

Masses were redetermined in three cases with the fit procedure detailed in \citet{ser+al14b}. Briefly, the observed shear profile is fitted to a spherical NFW functional through the function,
\beq
\label{eq_chi_WL}
\chi_\mathrm{WL}^2 ( M_{200},c_{200} )=\sum_i \left[ \frac{g_{+}(\theta_i)-g_{+}^\mathrm{NFW}(\theta_i; M_{200},c_{200})}{\delta_{+}(\theta_i)}\right]^2,
\eeq
where $g_{+}$  is the reduced tangential shear at angular position $\theta$ and $\delta_{+}$ is the observational uncertainty. 

When a strong lensing constraint was available, the effective angular Einstein radius $\theta_\mathrm{E}$ was fitted through
\beq
\label{eq_chi_SL}
\chi_\mathrm{SL}^2 ( M_{200},c_{200} )=
\left[\frac{
\theta_\mathrm{E}-\theta_\mathrm{E}^\mathrm{NFW}(M_{200},c_{200})}{\delta \theta_\mathrm{E}}
\right]^2.
\eeq
Expressions for the lensing quantities of the NFW halo can be found in \citet{bar96,wr+br00}. The total likelihood is ${\cal L} \propto \exp\{-(\chi^2_\mathrm{WL}+\chi^2_\mathrm{SL})/2\}$. For the catalog, I considered uniform priors in the ranges $0.02 \le M_{200}/(10^{14}h^{-1}M_\odot) \le 100$ and $0.02 \le c_{200} \le 20$. The parameters and their uncertainties were finally derived as the bi-weight estimators of the marginalised posterior probability densities.

For the Local Cluster Substructure Survey (LOCUSS)  sample in \citet{oka+al10}, I fitted the published shear profiles in order to derive the masses of all the 30 clusters of the sample, rather than the 26 reported in \citet[ table 6]{oka+al10}. Shear measurements in \citet{oka+al10} are biased low due to contamination effects and systematics in shape measurements \citep{oka+al13}. We then corrected the fitted masses according to the factors reported in \citet[ table 2]{oka+al13}.

I also refitted the clusters previously analysed in \citet{se+co13}. The fit procedure was slightly improved since, see \citet{ser+al14b}. For the catalog, I used the updated mass determinations.

Finally, \citet{mah+al07} published the shear profile of ABELL 478 but they did not report the mass determination. Values listed in the catalogs are the result of the fit procedure I performed.


\subsection{Intentional omissions}

There was a number of intentional omissions. I required that each lensing cluster was confirmed by independent observations. Lensing peaks without an optical, X-ray or SZ counterpart were excised from the catalog. This may be the case of some weak-lensing shear-selected halos or lensing peaks found in pilot programs targeting fields centred on active galactic nuclei or quasars \citep{wol+al02}.

I did not include some lensing analyses of single clusters that were later refined/improved by the same authors or collaboration. Just as an example, this is the case of the analyses of the high redshift clusters in \citet{jee+al05a,jee+al05b}, \citet{jee+al06}, and \citet{je+ty09} that were later revised in \citet{jee+al11}.

I considered only lensing studies performed under the assumption of spherical symmetry. Unfortunately, there is just a handful of clusters with triaxial analyses \citep[][ and references therein]{ogu+al05,cor+al09,se+um11,ser+al13,mor+al12,lim+al13}. For homogeneity reasons, I excluded them.

Complex cluster morphologies may be separated in multiple peaks by high resolution WL analyses. To compile the catalog with unique entries, LC$^2$-{\it single}, I only considered masses measured with a single halo analysis. Masses of substructures and multiple peaks associated to the same clusters are reported in LC$^2$-{\it substructure}.

\subsection{Cluster identifications}

The same cluster may appear in several analyses under different names and with different quoted redshifts and locations. To standardise the notation, I reported the NASA/IPAC Extragalactic Database\footnote{\url{http://ned.ipac.caltech.edu/}.} (NED) preferred name and the NED's coordinates and redshift for each object. Most of the clusters were identified by name. A few of them were associated by matching positions. 

Since most of the lenses which were not associated by name in NED are secondary halos in merging or complex systems, or shear-selected peaks found in dense fields, I could not adopt a fixed search radius when cross-checking with the NED. In fact, a blind matching based on a fixed aperture can associate the same NED counterpart to multiple, separate lenses, which we know to be distinct according to the reference paper. The association by position was then performed cluster-by-cluster. A limited number of lenses, mostly SZ or shear-selected halos, lacked the NED identification.

Control of repeated entries was performed by looking for repeated NED associations. For clusters which were not identified by querying the NED, I also looked for matches of both position and redshift. If the cluster's coordinates were missing in the original papers, I used the location obtained from querying the NED.

\section{Catalog presentation}
\label{sec_pres}

\begin{table*}
\caption{The first 50 entries of the LC$^2$-single catalogue. The full catalogs are available in electronic form. Columns are described in Section~\ref{sec_summ}.}
\label{tab_cata}
\centering
\rotatebox{90}{
\resizebox{23cm}{!}{\
\begin{tabular}{ l l l l l l l l l l l r r r r r r r r }      
\hline
\hline
 Name  & Right Ascension  &  Declination &  $z$ &  Match &  NED's primary name &  NED's RA &  NED's DEC &  NED's $z$ &  Autor Code &  ADS's bibcode &  $M_{2500}$ &  $\delta M_{2500}$ &   $M_{500}$ &  $\delta M_{500}$ &   $M_{200}$ &  $\delta M_{200}$ &   $M_\mathrm{vir}$ &  $\delta M_\mathrm{vir}$ \\
  & (J2000) &  (J2000) &   &  &   &  (J2000) &  (J2000) &   &  &  &  $[10^{14}M_\odot]$ &  $[10^{14}M_\odot]$ &   $[10^{14}M_\odot]$ &  $[10^{14}M_\odot]$ &   $[10^{14}M_\odot]$ &  $[10^{14}M_\odot]$ &   $[10^{14}M_\odot]$ &  $[10^{14}M_\odot]$ \\
(1-2) & (3) & (4) & (5) & (6) & (7-8) & (9) & 10) & (11) & (12) & (13) & (14) & (15) & (16) & (17) & (18) & (19) & (20)  & (21) \\
 \hline
ABELL 2744                    & 00:14:20.67  & --30:24:00.86 & 0.308   & N  & ABELL 2744                    & 00:14:18.90 & --30:23:22.0 & 0.308    & merten+11            & 2011MNRAS.417..333M   &  4.962 &  1.597 & 15.412 &  4.959 & 24.054 &  7.741 & 29.025 &  9.340 \\
CL 0016+16                    & 00:18:33.445 & +16:26:13.00 & 0.547   & N  & Cl 0016+16                    & 00:18:33.84 & +16:26:16.6 & 0.541    & applegate+14         & 2014MNRAS.439...48A   &  7.122 &  1.842 & 17.821 &  4.609 & 25.756 &  6.661 & 29.101 &  7.526 \\
ABELL 22                      & 00:20:38.6   & --25:43:19.2  & 0.141   & N  & ABELL 22                      & 00:20:42.80 & --25:42:37.0 & 0.142352 & cypriano+04          & 2004ApJ...613...95C   &  1.012 &  0.512 &  2.263 &  1.146 &  3.578 &  1.812 &  4.760 &  2.410 \\
ACT-CL J0022.2-0036           & 00:22:13.0   & --00:36:33.8  & 0.805   & NA & NA NA                         & NA          & NA          & NA       & miyatake+13          & 2013MNRAS.429.3627M   &  2.824 &  1.456 &  7.860 &  4.053 & 11.798 &  6.083 & 13.118 &  6.764 \\
MACS J0025.4-1222             & 00:25:29.907 & --12:22:44.64 & 0.585   & N  & MACS J0025.4-1222             & 00:25:29.38 & --12:22:37.1 & 0.5843   & applegate+14         & 2014MNRAS.439...48A   &  3.965 &  1.392 &  9.922 &  3.483 & 14.340 &  5.034 & 16.130 &  5.663 \\
Cl 0024+17                    & 00:26:36.0   & +17:08:36    & 0.39    & N  & ZwCl 0024.0+1652              & 00:26:35.70 & +17:09:46.0 & 0.39     & umetsu+11            & 2011ApJ...729..127U   &  6.600 &  0.900 & 12.329 &  1.943 & 17.057 &  2.686 & 18.986 &  3.200 \\
CL 0030+2618                  & 00:30:33.6   & +26:18:16    & 0.5     & N  & WARP J0030.5+2618             & 00:30:33.20 & +26:18:19.0 & 0.5      & israel+12            & 2012A\%26A...546A..79I &  1.803 &  0.516 &  4.512 &  1.292 &  6.521 &  1.867 &  7.412 &  2.122 \\
ABELL 68                      & 00:37:05.947 & +09:09:36.02 & 0.255   & N  & ABELL 68                      & 00:37:05.30 & +09:09:11.0 & 0.255    & applegate+14         & 2014MNRAS.439...48A   &  3.665 &  0.634 &  9.171 &  1.587 & 13.254 &  2.294 & 15.670 &  2.712 \\
ABELL 85                      & 00:41:48.7   & --09:19:04.8  & 0.056   & N  & ABELL 85                      & 00:41:50.10 & --09:18:07.0 & 0.055061 & cypriano+04          & 2004ApJ...613...95C   &  2.048 &  0.557 &  4.579 &  1.245 &  7.240 &  1.968 &  9.947 &  2.704 \\
ABELL 2811                    & 00:42:07.9   & --28:32:09.6  & 0.108   & N  & ABELL 2811                    & 00:42:08.70 & --28:32:09.0 & 0.107908 & cypriano+04          & 2004ApJ...613...95C   &  1.767 &  0.849 &  3.952 &  1.898 &  6.248 &  3.000 &  8.414 &  4.040 \\
ABELL 115                     & 00:55:59.8   & +26:22:40.8  & 0.1971  & N  & ABELL 115                     & 00:55:59.80 & +26:22:41.0 & 0.1971   & okabe+10             & 2010PASJ...62..811O   &  1.478 &  0.936 &  4.832 &  3.061 &  7.427 &  4.705 &  9.108 &  5.771 \\
Cl 0054-27                    & 00:56:37.4   & --27:30:47    & 0.56    & N  & ABELL 2843                    & 00:56:37.38 & --27:30:46.7 & 0.56     & smail+97             & 1997ApJ...479...70S   &  1.131 &  0.619 &  3.391 &  1.857 &  5.227 &  2.862 &  6.012 &  3.292 \\
RX J0056.9-2740               & 00:56:56.98  & --27:40:29.9  & 0.563   & N  & Cl 0054-2756                  & 00:56:56.87 & --27:40:30.1 & 0.56     & hoekstra+11          & 2011ApJ...726...48H   &  0.640 &  0.360 &  1.862 &  1.048 &  2.840 &  1.597 &  3.255 &  1.831 \\
ACT-CL J0102-4915             & 01:02:52.50  & --49:14:58.0  & 0.87    & N  & SPT-CL J0102-4915             & 01:02:53.00 & --49:15:19.0 & 0.75     & jee+14               & 2014ApJ...785...20J   &  4.391 &  0.847 & 16.800 &  3.200 & 25.400 &  4.900 & 28.661 &  5.529 \\
ABELL 141                     & 01:05:36.3   & --24:39:20    & 0.23    & N  & ABELL 141                     & 01:05:34.80 & --24:39:17.0 & 0.23     & dahle+02             & 2002ApJS..139..313D   &  3.225 &  1.301 &  7.212 &  2.909 & 11.403 &  4.599 & 14.723 &  5.939 \\
ZwCl 0104.4+0048              & 01:06:48.5   & +01:02:42.0  & 0.2545  & N  & ZwCl 0104.4+0048              & 01:06:58.07 & +01:04:01.4 & 0.2545   & okabe+10             & 2010PASJ...62..811O   &  1.008 &  0.262 &  1.841 &  0.478 &  2.299 &  0.597 &  2.546 &  0.661 \\
RX J0110.3+1938               & 01:10:18.22  & +19:38:19.4  & 0.317   & N  & WARP J0110.3+1938             & 01:10:18.00 & +19:38:23.0 & 0.317    & hoekstra+11          & 2011ApJ...726...48H   &  0.580 &  0.300 &  1.585 &  0.820 &  2.363 &  1.222 &  2.796 &  1.446 \\
ABELL 209                     & 01:31:52.54  & --13:36:40.4  & 0.206   & N  & ABELL 209                     & 01:31:53.00 & --13:36:34.0 & 0.206    & umetsu+14            & 2014arXiv1404.1375U   &  4.033 &  0.688 & 11.573 &  1.796 & 17.559 &  2.993 & 21.473 &  3.922 \\
ABELL 222                     & 01:37:34.0   & --12:59:29    & 0.213   & N  & ABELL 222                     & 01:37:29.20 & --12:59:10.0 & 0.213    & mahdavi+13           & 2013ApJ...767..116M   &  1.572 &  0.555 &  5.649 &  1.246 &  8.559 &  1.888 & 10.417 &  2.298 \\
ABELL 223S                    & 01:37:56.0   & --12:49:10    & 0.207   & NA & NA NA                         & NA          & NA          & NA       & mahdavi+13           & 2013ApJ...767..116M   &  1.003 &  0.482 &  6.863 &  1.936 & 10.435 &  2.943 & 12.738 &  3.593 \\
ABELL 223N                    & 01:38:02.3   & --12:45:20    & 0.207   & NA & NA NA                         & NA          & NA          & NA       & hoekstra+12          & 2012MNRAS.427.1298H   &  2.100 &  0.550 &  5.500 &  1.900 &  8.320 &  2.142 & 10.100 &  2.600 \\
RX J0142.0+2131               & 01:42:03.311 & +21:31:22.64 & 0.28    & N  & MCXC J0142.0+2131             & 01:42:02.60 & +21:31:19.0 & 0.2803   & applegate+14         & 2014MNRAS.439...48A   &  1.847 &  0.743 &  4.621 &  1.860 &  6.678 &  2.688 &  7.858 &  3.162 \\
CL J0152-1357                 & 01:52:41.0   & --13:57:45    & 0.84    & N  & WARP J0152.7-1357             & 01:52:41.00 & --13:57:45.0 & 0.831    & sereno\&13            & 2013MNRAS.434..878S   &  1.382 &  0.291 &  2.259 &  0.475 &  2.800 &  0.589 &  2.964 &  0.624 \\
ABELL 267                     & 01:52:42.0   & +01:00:26    & 0.23    & N  & ABELL 267                     & 01:52:52.26 & +01:02:45.8 & 0.231    & mahdavi+13           & 2013ApJ...767..116M   &  2.053 &  0.428 &  5.245 &  1.523 &  7.948 &  2.308 &  9.637 &  2.798 \\
RX J0154.2-5937               & 01:54:13.72  & --59:37:31.0  & 0.36    & N  & 400d J0154-5937               & 01:54:14.80 & --59:37:48.0 & 0.36     & hoekstra+11          & 2011ApJ...726...48H   &  0.340 &  0.200 &  0.910 &  0.535 &  1.348 &  0.793 &  1.578 &  0.928 \\
CL 0159+0030                  & 01:59:18.2   & +00:30:09    & 0.39    & N  & NSCS J015924+003024           & 01:59:17.00 & +00:30:10.4 & 0.386    & israel+12            & 2012A\%26A...546A..79I &  1.443 &  0.634 &  3.466 &  1.524 &  4.937 &  2.170 &  5.671 &  2.493 \\
CFHTLS c12-w1                 & 02:01:18.00  & --07:39:03.60 & 0.14    & NA & NA NA                         & NA          & NA          & NA       & shan+12              & 2012ApJ...748...56S   &  0.107 &  0.023 &  0.259 &  0.055 &  0.371 &  0.078 &  0.447 &  0.095 \\
CFHTLS c3-w1                  & 02:01:41.04  & --05:01:48.00 & 0.28    & NA & NA NA                         & NA          & NA          & NA       & shan+12              & 2012ApJ...748...56S   &  0.987 &  0.183 &  2.813 &  0.522 &  4.257 &  0.790 &  5.104 &  0.947 \\
ABELL 291                     & 02:01:44.2   & --02:12:03.0  & 0.196   & N  & ABELL 291                     & 02:01:44.20 & --02:12:03.0 & 0.197    & okabe+10             & 2010PASJ...62..811O   &  1.166 &  0.345 &  5.222 &  1.545 &  9.022 &  2.669 & 11.658 &  3.448 \\
CFHTLS c16-w1                 & 02:01:48.48  & --10:51:10.80 & 0.66    & NA & NA NA                         & NA          & NA          & NA       & shan+12              & 2012ApJ...748...56S   &  0.617 &  0.094 &  1.870 &  0.284 &  2.892 &  0.440 &  3.289 &  0.500 \\
CFHTLS c14-w1                 & 02:02:08.16  & --08:26:13.20 & 0.4     & P  & GMBCG J030.53418-08.43703     & 02:02:08.20 & --08:26:13.3 & 0.349    & shan+12              & 2012ApJ...748...56S   &  0.458 &  0.054 &  1.285 &  0.151 &  1.934 &  0.227 &  2.262 &  0.266 \\
CFHTLS c2-w1                  & 02:02:30.00  & --03:57:57.60 & 0.48    & NA & NA NA                         & NA          & NA          & NA       & shan+12              & 2012ApJ...748...56S   &  4.884 &  2.826 & 16.261 &  9.408 & 26.022 & 15.057 & 30.663 & 17.742 \\
CFHTLS c13-w1                 & 02:02:49.92  & --09:20:13.20 & 0.23    & P  & ABELL 298                     & 02:02:48.90 & --09:19:57.0 & 0.14315  & shan+12              & 2012ApJ...748...56S   &  0.054 &  0.031 &  0.130 &  0.075 &  0.185 &  0.107 &  0.219 &  0.126 \\
CFHTLS c6-w1                  & 02:03:21.60  & --07:19:51.60 & 0.27    & NA & NA NA                         & NA          & NA          & NA       & shan+12              & 2012ApJ...748...56S   &  0.258 &  0.057 &  0.678 &  0.150 &  0.997 &  0.221 &  1.184 &  0.262 \\
CFHTLS c5-w1                  & 02:03:26.40  & --05:55:30.00 & 0.38    & NA & NA NA                         & NA          & NA          & NA       & shan+12              & 2012ApJ...748...56S   &  0.246 &  0.052 &  0.662 &  0.139 &  0.981 &  0.206 &  1.146 &  0.241 \\
CFHTLS c15-w1                 & 02:03:28.80  & --09:49:01.20 & 0.32    & P  & GMBCG J030.86962-09.81667     & 02:03:28.71 & --09:49:00.0 & 0.33     & shan+12              & 2012ApJ...748...56S   &  0.220 &  0.055 &  0.580 &  0.145 &  0.854 &  0.213 &  1.005 &  0.251 \\
CFHTLS c9-w1                  & 02:03:32.16  & --06:44:09.60 & 0.81    & NA & NA NA                         & NA          & NA          & NA       & shan+12              & 2012ApJ...748...56S   &  1.042 &  0.261 &  3.371 &  0.845 &  5.339 &  1.338 &  6.012 &  1.507 \\
CFHTLS c1-w1                  & 02:03:57.60  & --04:13:22.80 & 0.18    & NA & NA NA                         & NA          & NA          & NA       & shan+12              & 2012ApJ...748...56S   &  1.956 &  1.176 &  5.652 &  3.398 &  8.597 &  5.168 & 10.562 &  6.350 \\
CFHTLS c7-w1                  & 02:04:18.48  & --07:12:46.80 & 0.33    & NA & NA NA                         & NA          & NA          & NA       & shan+12              & 2012ApJ...748...56S   &  0.682 &  0.486 &  1.924 &  1.372 &  2.903 &  2.070 &  3.442 &  2.455 \\
CFHTLS c22-w1                 & 02:05:05.28  & --05:55:22.80 & 0.41    & NA & NA NA                         & NA          & NA          & NA       & shan+12              & 2012ApJ...748...56S   &  0.088 &  0.055 &  0.224 &  0.141 &  0.327 &  0.206 &  0.377 &  0.238 \\
CFHTLS c26-w1                 & 02:05:16.80  & --07:40:48.00 & 0.24    & NA & NA NA                         & NA          & NA          & NA       & shan+12              & 2012ApJ...748...56S   &  0.169 &  0.056 &  0.431 &  0.142 &  0.627 &  0.207 &  0.745 &  0.246 \\
CFHTLS c24-w1                 & 02:05:25.92  & --07:35:16.80 & 0.4     & P  & CFHTLS:[DAC2011] W1-0493      & 02:05:25.92 & --07:35:22.6 & 0.47     & shan+12              & 2012ApJ...748...56S   &  0.455 &  0.185 &  1.275 &  0.520 &  1.919 &  0.782 &  2.245 &  0.915 \\
CFHTLS c27-w1                 & 02:06:22.80  & --08:48:25.20 & 0.26    & NA & NA NA                         & NA          & NA          & NA       & shan+12              & 2012ApJ...748...56S   &  0.653 &  0.414 &  1.807 &  1.146 &  2.706 &  1.716 &  3.244 &  2.057 \\
CFHTLS c28-w1                 & 02:07:04.56  & --08:29:42.00 & 0.08    & NA & NA NA                         & NA          & NA          & NA       & shan+12              & 2012ApJ...748...56S   &  0.230 &  0.066 &  0.574 &  0.163 &  0.828 &  0.236 &  1.016 &  0.289 \\
CFHTLS c18-w1                 & 02:07:11.28  & --04:00:07.20 & 0.25    & NA & NA NA                         & NA          & NA          & NA       & shan+12              & 2012ApJ...748...56S   &  2.256 &  1.374 &  6.711 &  4.087 & 10.313 &  6.282 & 12.523 &  7.628 \\
CFHTLS c17-w1                 & 02:08:05.04  & --04:34:58.80 & 0.32    & NA & NA NA                         & NA          & NA          & NA       & shan+12              & 2012ApJ...748...56S   &  2.035 &  1.135 &  6.130 &  3.420 &  9.464 &  5.280 & 11.343 &  6.328 \\
CFHTLS c39-w1                 & 02:08:32.40  & --07:43:48.00 & 0.34    & NA & NA NA                         & NA          & NA          & NA       & shan+12              & 2012ApJ...748...56S   &  0.817 &  0.528 &  2.337 &  1.511 &  3.543 &  2.291 &  4.200 &  2.716 \\
CFHTLS c42-w1                 & 02:09:15.12  & --09:14:38.40 & 0.71    & NA & NA NA                         & NA          & NA          & NA       & shan+12              & 2012ApJ...748...56S   &  0.382 &  0.148 &  1.135 &  0.438 &  1.743 &  0.673 &  1.967 &  0.760 \\
CFHTLS c33-w1                 & 02:09:44.40  & --05:43:48.00 & 0.23    & NA & NA NA                         & NA          & NA          & NA       & shan+12              & 2012ApJ...748...56S   &  0.340 &  0.038 &  0.899 &  0.101 &  1.325 &  0.150 &  1.588 &  0.179 \\
ABELL 315                     & 02:10:03.0   & --00:59:52    & 0.1754  & N  & ABELL 315                     & 02:10:03.06 & --00:59:51.8 & 0.1754   & dietrich+09          & 2009A\%26A...499..669D &  1.172 &  0.331 &  3.215 &  0.909 &  4.800 &  1.357 &  5.854 &  1.655 \\
\hline
\end{tabular}
}
}
\end{table*}



I compiled three catalogs. The LC$^2$-{\it single} lists all clusters and groups whose mass was determined with a single-halo modelling, no matter what the dynamical state, and contains virtually no multiple entries. 

The LC$^2$-{\it substructure} lists separately the main components and the secondary haloes of complex systems, whose masses were derived with a multiple-halo analysis. As for LC$^2$-{\it single}, duplicate entries were eliminated. There is some redundancy between LC$^2$-{\it single} and LC$^2$-{\it substructure}. Some systems may appear as a single halo in LC$^2$-{\it single} and as a main halo with substructures in LC$^2$-{\it substructure}.

LC$^2$-{\it all} comprises the full body of information I found and reduced from literature. Multiple entries are present, as well as single- or multiple-halo analyses of the same lens. The LC$^2$-{\it single} and LC$^2$-{\it substructure} are subsamples with unique entries of LC$^2$-{\it all}. When a cluster had multiple analyses available in literature, I picked for the LC$^2$-{\it single} either the most recent analysis or that based on deeper observations.
  
Table~\ref{tab_cata} presents an extract of LC$^2$-{\it single}, in terms of the first 50 entries. In each catalog, objects are ordered by right ascension. The format is as follows.
\begin{description}
\item Cols. 1-2: name of cluster as designated in the original lensing paper.
\item Cols. 3-4: right ascension RA (J2000) and declination DEC (J2000), as quoted in the original lensing paper. If coordinates are not quoted in the source paper or in a companion one, I reported the coordinates of the NED's association.
\item Col. 5: redshift $z$, as reported in the original lensing paper.
\item Col. 6: external validation through NED. `N': the NED's object was associated by name; `P': the NED's object was associated by positional matching; `NA': no found association.
\item Cols. 7-11: as in cols. 1-5, but for the NED's association.
\item Col. 12: author code.
\item Col. 13: ADS's bibliographic code.
\item Cols. 14-15: over-density mass $M_{2500}$ and related uncertainty $\delta M_{2500}$, in units of $10^{14}M_\odot$.
\item Cols. 16-17: as for cols. 14-15, but for the over-density mass $M_{500}$.
\item Cols. 18-19: as for cols. 14-15, but for the over-density mass $M_{200}$.
\item Cols. 20-21: as for cols. 14-15, but for the virial mass $M_\mathrm{vir}$.
\end{description}

\subsection{Basic properties}
\label{sec_basi}

\begin{figure}
       \resizebox{\hsize}{!}{\includegraphics{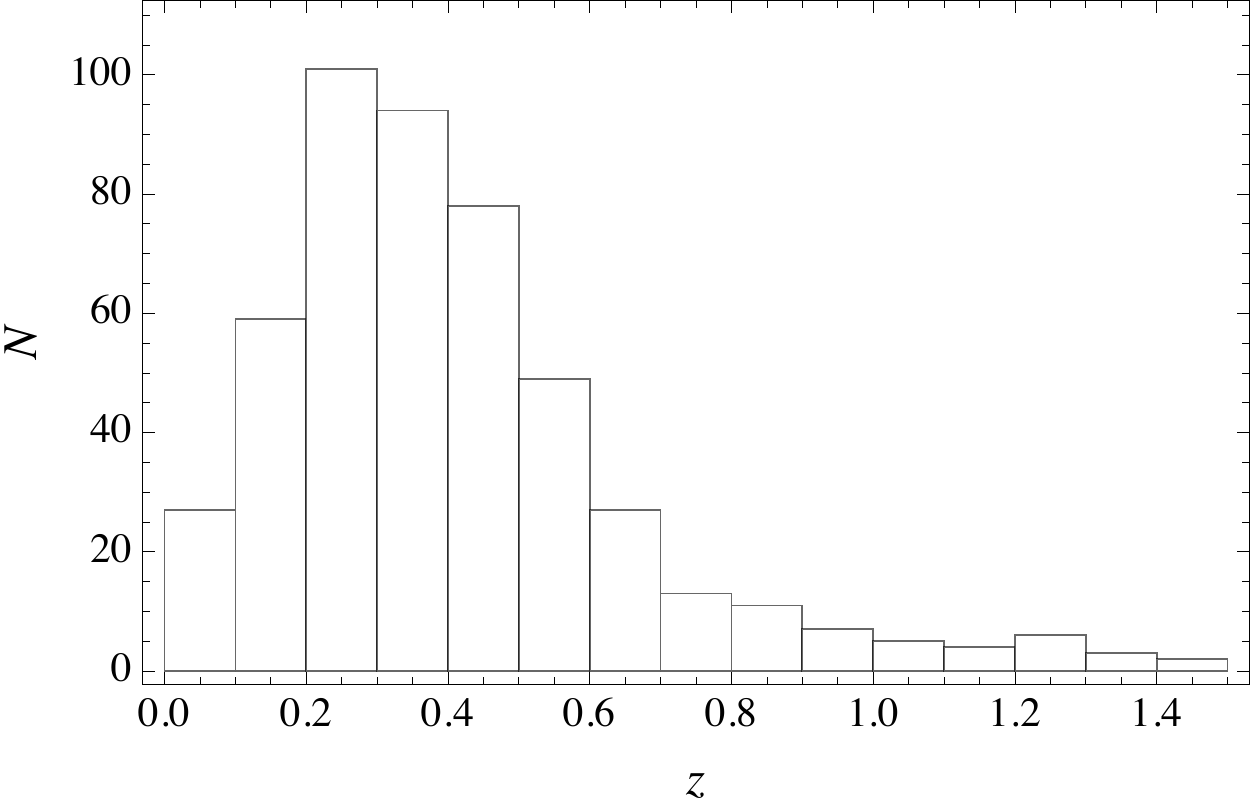}}
       \caption{Redshift distribution of the 485 WL clusters in the LC$^2$-{\it single} catalog.}
	\label{fig_histo_z}
\end{figure}

\begin{figure}
\begin{tabular}{c}
\resizebox{\hsize}{!}{\includegraphics{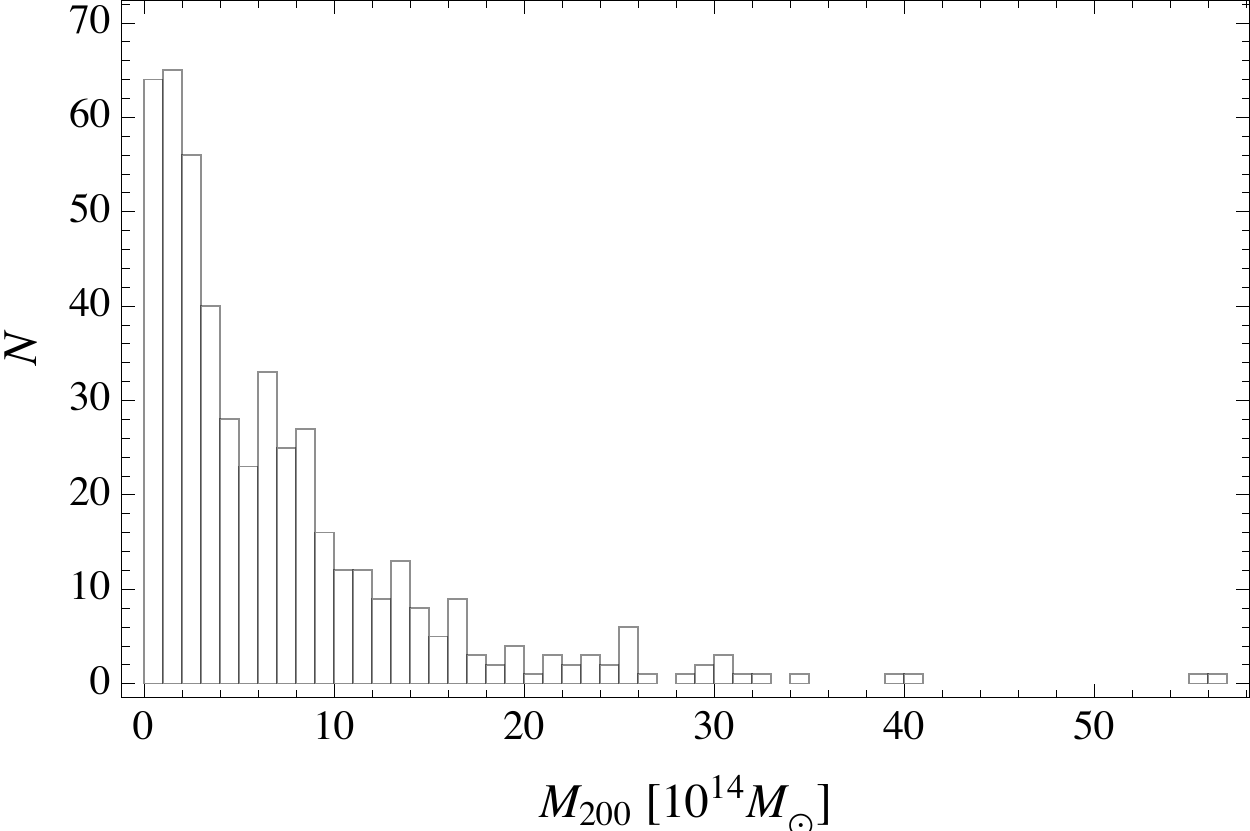}} \\
\noalign{\smallskip}  
\resizebox{\hsize}{!}{\includegraphics{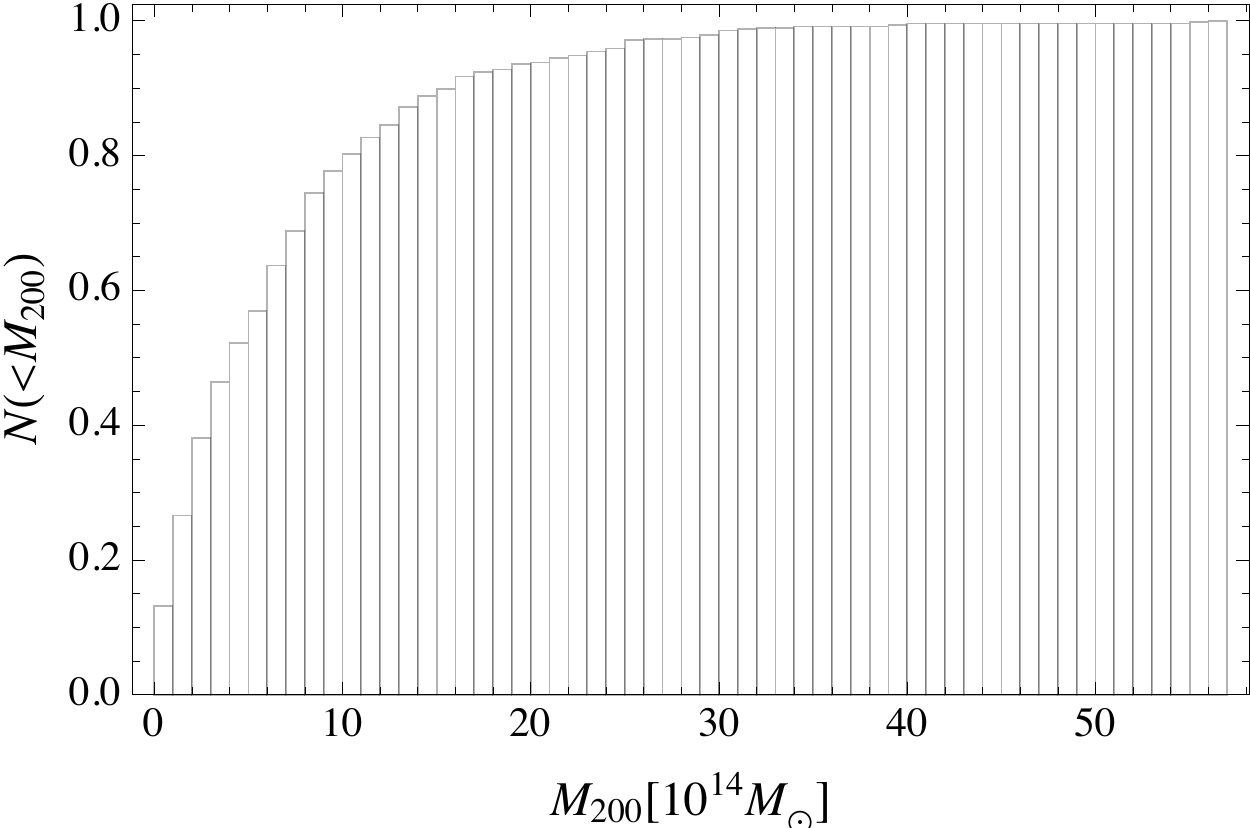}}
\end{tabular}
\caption{Mass distribution of the 485 WL clusters in the LC$^2$-{\it single} catalog. {\it Top panel}: histogram of the mass distribution. {\it Bottom panel}: normalised cumulative function. $M_{200}$ is in units of $10^{14}M_\odot$.}
\label{fig_histo_M500_WL}
\end{figure}

\begin{figure}
       \resizebox{\hsize}{!}{\includegraphics{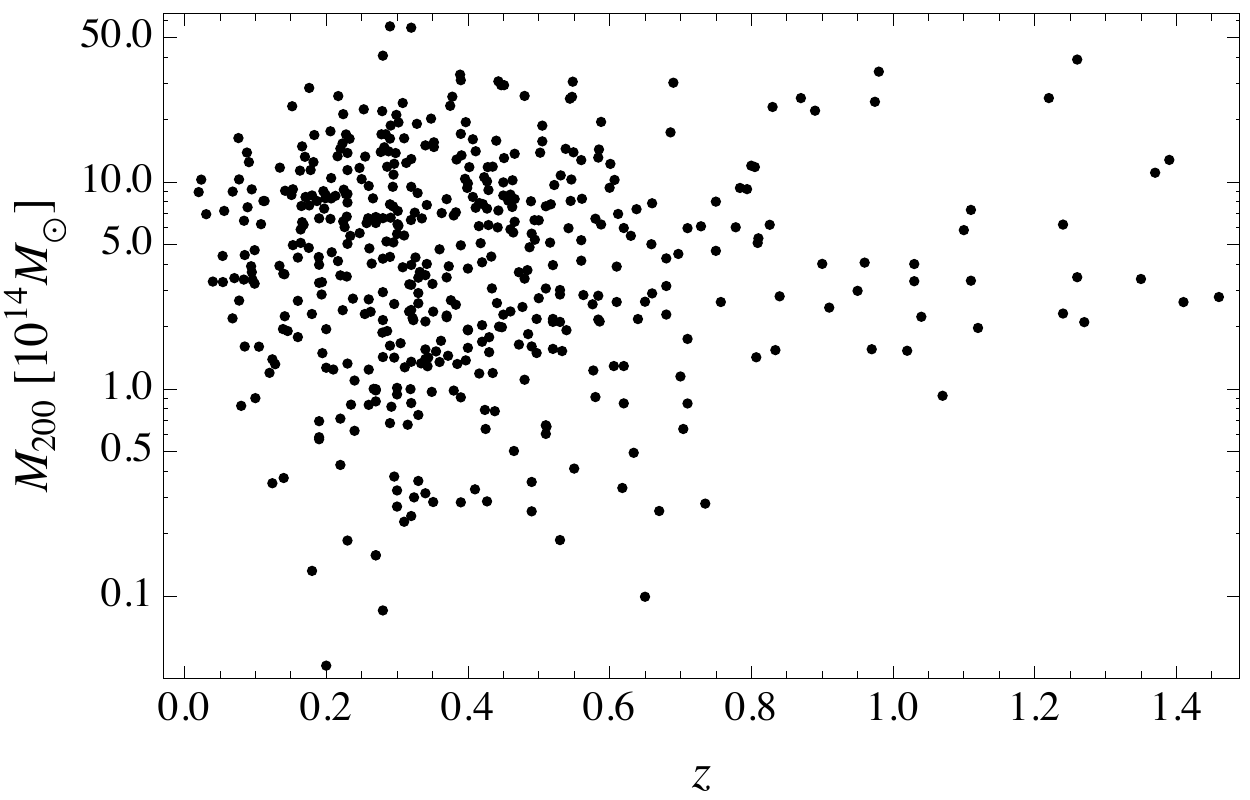}}
       \caption{Redshift versus mass for the 485 WL clusters in the LC$^2$-{\it single} catalog. $M_{200}$ is in units of $10^{14}M_\odot$.}
	\label{fig_z_M200}
\end{figure}

I discuss the basic properties of the collected clusters. 507 clusters, groups, or sub-structures were analysed in published lensing studies. 131 objects were studied by at least two independent groups. The most popular targets are ABELL 209, 1835, and 2261, with ten independent analyses each, and ABELL 611 and 1689 (9 analyses each). Overall, we found 822 mass determinations.

The {\it single} catalog contains 485 unique entries. The redshift distribution of the (unique) clusters (see Fig.~\ref{fig_histo_z}) has a large range, $0.02\ls z\ls 1.46$, with a peak at $z \sim 0.35$, where lensing studies are optimised. The tail at large redshift includes 50 (20) clusters at $z>0.7$ (1.0).

Weak lensing is better suited to measure massive clusters. The mass distribution has a median $M_{200} \sim 4.5 \times 10^{14} M_\odot$ and extends to $M_{200}$ larger than $5\times 10^{15} M_\odot$, see Fig.~\ref{fig_histo_M500_WL}. Shear or X-ray selected groups of clusters mostly populate the less massive bins.

Due to the heterogeneous nature of the sample there is no evident trend in cluster masses with redshift, see Fig.~\ref{fig_z_M200}. Approximated selection functions might be derived only for specific subsamples.

\section{Conclusions}
\label{sec_summ}

A standardised collection of weak lensing masses can be useful for X-ray, SZ, and other multi-wavelength studies. I compiled from literature three catalogues. The LC$^2$-{\it all}, -{\it single} and -{\it substructure} catalogs comprise 822, 485 and 18 groups and clusters, respectively. 

The LC$^2$-{\it all}  catalog is a repository of all the main information on clusters with measured lensing mass I found in literature. LC$^2$-{\it single} is a list of unique entries. LC$^2$-{\it substructure} focuses on complex structures.

The full catalogs are publicly available in electronic format\footnote{\url{http://pico.bo.astro.it/\textasciitilde sereno/CoMaLit/LC2/}.}. The first version of the catalogs is released together with this presentation paper. The catalogs will be periodically updated.

\section*{Acknowledgements}
I thank S.~Ettori, L.~Moscardini, and K.~Umetsu for fruitful discussions. I acknowledge financial contributions from contracts ASI/INAF n.I/023/12/0 `Attivit\`a relative alla fase B2/C per la missione Euclid', PRIN MIUR 2010-2011 `The dark Universe and the cosmic evolution of baryons: from current surveys to Euclid', and PRIN INAF 2012 `The Universe in the box: multiscale simulations of cosmic structure'. I thank NASA. This research has made use of NASA's Astrophysics Data System (ADS) and of the NASA/IPAC Extragalactic Database (NED), which is operated by the Jet Propulsion Laboratory, California Institute of Technology, under contract with the National Aeronautics and Space Administration.

\bibliographystyle{mn2e_fix_Williams}

\setlength{\bibhang}{2.0em}

\end{document}